\documentclass{aastex631}

\usepackage[utf8]{inputenc}
\usepackage{xcolor}
\usepackage{hyperref}
\usepackage{graphicx}
\usepackage{mathrsfs}

\newcommand{\myurl}[2][red]{\href{#2}{\color{#1}{#2}}}%

\begin{document}

\received{}
\revised{}
\accepted{}
\submitjournal{ApJS}

\shorttitle{PHOEBE IX: Spectroscopic module}
\shortauthors{Brož et al.}

\title{Physics Of Eclipsing Binaries. IX. Spectroscopic module}

\correspondingauthor{Miroslav Brož}
\email{mira@sirrah.troja.mff.cuni.cz}

\author[0000-0003-2763-1411]{Miroslav Brož}
\affiliation{Charles University, Faculty of Mathematics and Physics, Institute of Astronomy, V Holešovičkách 2, CZ-18200 Praha 8}

\author[0000-0002-1913-0281]{Andrej Pr\v sa}
\affiliation{Villanova University, Dept.~of Astrophysics and Planetary Sciences, 800 E.\ Lancaster Ave, Villanova, PA 19085, USA}

\author[0000-0002-5442-8550]{Kyle E.~Conroy}
\affiliation{Villanova University, Dept.~of Astrophysics and Planetary Sciences, 800 E.\ Lancaster Ave, Villanova, PA 19085, USA}

\author[0000-0001-6566-7568]{Michael Abdul-Masih}
\affiliation{Instituto de Astrofísica de Canarias, C. Vía Láctea, s/n, 38205 La Laguna, Santa Cruz de Tenerife, Spain}


\begin{abstract}
Spectroscopic observations constrain the fundamental properties of stellar atmospheres,
in particular,
the effective temperature,
the gravitational acceleration, or
the metallicity.
In this work,
we describe the spectroscopic module for Phoebe,
which allows for modelling of spectra,
either normalized, or in absolute units (${\rm W}\,{\rm m}^{-2}\,{\rm m}^{-1}$).
The module is based on extensive grids of synthetic spectra,
taken from literature,
which are interpolated and integrated over the surface.
As an approximation,
we assume that limb darkening is given by an analytical law,
while other effects (e.g., eclipses) are treated self-consistently.
Our approach is suitable for
single stars, binaries, or multiples,
and can be further extended to systems with pulsating components.
\textcolor{red}{This draft refers to a development version of Phoebe, available at \myurl{https://github.com/miroslavbroz/phoebe2/tree/spectroscopy2}.
It is not yet included in the official Phoebe repository!}
\end{abstract}


\section{Introduction}

Models of eclipsing binaries are commonly constrained
by two (or more) types of observations,
e.g., by light curves and radial velocities,
in order to obtain the absolute values of parameters
(masses, radii, temperatures, orbit, or distance).
A number of correlations among these parameters exist
\citep{Conroy_2020ApJS..250...34C},
which could, in principle,
significantly decrease the uncertainties of parameters
(see, e.g., \citealt{Suzuki_2012ApJ...746...85S}).
In particular,
the gravitational acceleration at a stellar surface
\begin{equation}
g = {Gm\over R^2}
\end{equation}
is certainly correlated with
the mass $m$,
and the radius $R$;
the bolometric flux
\begin{equation}
\Phi = \left({R\over d}\right)^2\sigma T_{\rm eff}^4\,,
\end{equation}
again,
with the radius $R$,
the distance $d$,
and the effective temperature $T_{\rm eff}$;
or more specifically,
such relations also exist for individual passbands.
Last but not least,
chemical composition,
internal structure, or
pulsation frequencies,
are closely related to the metallicity $\mathscr{Z}$.
All these parameters could be constrained by spectroscopy.

A stellar atmosphere and its emergent spectrum is therefore parameterized by
$T_{\rm eff}$, $\log g$, $\mathscr{Z}$,
and possibly also by the direction
$\mu \equiv \cos\theta$,
if a full field is available
\citep{Puls_2005A&A...435..669P,Abdul_2020A&A...636A..59A,Abdul_2023A&A...669L..11A}.
One can use existing tools for interpolation in grids of synthetic spectra,
e.g., Pyterpol,
previously used for $\xi$~Tau
\citep{Nemravova_2016A&A...594A..55N,Broz_2017ApJS..230...19B}
and compare synthetic to observed spectra.
However, it is still important to combine spectra with other datasets,
otherwise the model would remain unconstrained, or poorly constrained.

In this work,
we thus implement a spectroscopic module in Phoebe
\citep{Prsa_2016ApJS..227...29P,Horvat_2018ApJS..237...26H,Jones_2020ApJS..247...63J,Conroy_2020ApJS..250...34C,Abdul_2020A&A...636A..59A}
to improve modelling of binaries (or multiple systems).
Apart from
light curves and radial velocities,
normalized (rectified) spectra
and/or spectral-energy distributions (SED)
could be used.
Ideally,
radial velocities should be entirely superseded by spectra,
which encompass the same information.


\section{Methods}\label{methods}

Single stars, binaries, or triples in Phoebe are all described
by means of triangular meshes
\citep{Prsa_2016ApJS..227...29P},
which closely follow the Roche potential
\citep{Roche_1873,Kopal_1959cbs..book.....K,Horvat_2018ApJS..237...26H}.
Summing over triangles belonging to one component
allows to compute also per-component properties.
We thus have two options for the spectroscopic module,
(i)~complex, with integration of spectra over triangles;
(ii)~simplified, with just one spectrum per component.
The former is much more precise,
but the latter is fast.

\subsection{Complex model}\label{complex}

In our complex model,
normalized synthetic spectra are generated for each triangle,
\begin{equation}
I_{\lambda,i} = \verb|sg.get_synthetic_spectrum|(T_{{\rm eff},i}, \log g_i, \mathscr{Z}_i) \quad\hbox{for }\forall i\,,
\end{equation}
parametrized by
$T_{\rm eff}$, the local effective temperature,
$\log g$, the local gravitational acceleration (in cgs units), and
$\mathscr{Z} = 10^{\rm abun}$, the component metallicity.
Typically, more than 1000 spectra per component are needed.
We note that the interpolation is not done in $\mu$
and that the limb darkening is discussed elsewhere
(see $I_{\rm pass}$ below).

The Doppler effect is accounted for in a standard way, as
\begin{equation}
\lambda' = \lambda\left(1+{v_{\rm rad}\over c}\right) \label{lambda_}
\end{equation}
where
$v_{\rm rad}$ is the radial velocity,
$c$ the speed of light.
In our specific case, we apply neither instrumental broadening,
since our synthetic spectra were broadened (to 0.01\,\AA),
nor rotational broadening,
since it is already accounted for in the radial motion of all triangles;
$I_\lambda' = I_\lambda$.%
\footnote{Optionally, one may apply rotational broadening (Eq.~(\ref{I_lambda_})) to each triangle,
according to its {\em differential\/} radial velocity of 3 vertices.}
Nevertheless, we always perform
a piece-wise linear interpolation to the observed wavelengths~($\lambda''$)
\begin{equation}
I_\lambda'' = \verb|pyterpolmu.interpolate_spectrum|(\lambda', I_\lambda', \lambda'')\,.\label{I_lambda__}
\end{equation}

Finally, the normalized monochromatic flux is computed as
\begin{equation}
\Phi_\lambda = {1\over L_{\rm tot}}\sum_i I_{{\rm pass},i} S_i \cos\theta_i f_i I_{\lambda,i}''\,,
\end{equation}
where the sum is over triangles,
$I_{{\rm pass},i}$ are the corresponding passband intensities,
which are limb-darkened and gravity-darkened
(i.e., the same as for light curve computations),
$S_i$ the surface areas,
$\cos\theta_i$ the cosines with respect to the local normals, and
$f_i$ fractions of triangles, which are visible.

\paragraph{SED}

When computing SEDs, we must use a 2nd grid of {\em absolute\/} synthetic spectra,
at stellar surface,
in ${\rm erg}\,{\rm s}^{-1}\,{\rm cm}^{-2}\,{\rm\AA}^{-1}$ units,
\begin{equation}
I_{\lambda,i} = \verb|sg2.get_synthetic_spectrum|(T_{{\rm eff},i}, \log g_i, \mathscr{Z}_i) \quad\hbox{for }\forall i\,.\label{I_lambda_sg2}
\end{equation}
The monochromatic flux,
at Earth,
in ${\rm W}\,{\rm m}^{-2}\,{\rm m}^{-1}$ units,
is computed as
\begin{equation}
\Phi_\lambda = {1\over\pi d^2} \sum_i {\rm lds}_i S_i \cos\theta_i f_i I_{\lambda,i}''\,,
\end{equation}
where
$d$ is the distance to the system,
${\rm lds}_i$ the limb darkening function values;
the factor of $\pi$ in the denominator is due to the surface area
(cancelling with $\pi R^2$),
and an additional factor of $10^7$ is due to units conversion.

Since we use our own synthetic spectra
(i.e., not the atmospheres included in Phoebe)
we have to account for the limb darkening
by choosing an analytical limb-darkening law
(linear, logarithmic, square-root, quadratic, or non-linear);
the limb darkening coefficients
should correspond to the range of $\lambda$,
of course.


\subsection{Simplified model}\label{simplified}

In our simplified model,
we estimate component luminosities and use them as weights for component spectra.
For simplicity, we compute the Planck function
\begin{equation}
B_\lambda = {2hc^2\over\lambda^5}{1\over \exp\bigl({hc\over\lambda kT_{\rm eff}}\bigr)-1}
\end{equation}
and the monochromatic luminosity,
\begin{equation}
L_\lambda = \pi R^2 B_\lambda\,.
\end{equation}
The respective synthetic spectrum, for $\mu = 1$, is obtained as
\begin{equation}
I_\lambda = \verb|sg.get_synthetic_spectrum|(T_{{\rm eff}}, \log g, \mathscr{Z})\,.
\end{equation}

The Doppler effect is computed as before (Eq.~(\ref{lambda_})).
We do not apply the instrumental broadening,
but do apply the rotational broadening,
\begin{equation}
I_\lambda' = 1-{\cal F}^{-1}\left[{\cal F}(1-I_\lambda)\,{\cal F}(K))\right]\,,\label{I_lambda_}
\end{equation}
where the fast Fourier transform is used (instead of convolution).
The respective kernel
\citep{Diaz_2011A&A...531A.143D}
\begin{equation}
K = {2\over\pi}{(1-\epsilon)\sqrt{\rm arg} + {\pi\over 2}\epsilon\,{\rm arg}\over v_{\rm rot}/c\,(1-{1\over 3}\epsilon)}\,,
\end{equation}
where $v_{\rm rot} \equiv \Omega R\sin i_\star$ is the projected rotation velocity,
${\rm arg} \equiv 1-\left({v/v_{\rm rot}}\right)^2$,
$\epsilon$ the linear limb-darkening coefficient;
the wavelength scale is equidistant in $\log\lambda$.
The kernel normalisation is done {\em ex-post\/}.
The interpolation to the observed wavelengths is as before (Eq.~(\ref{I_lambda__})).

Finally, the normalized monochromatic flux is computed as
\begin{equation}
\Phi_\lambda = {1\over L_{\rm tot}}\sum_i L_{\lambda,i} I_{\lambda,i}''\,,
\end{equation}
where the sum is over components.

\paragraph{SED}

Likewise, we obtain absolute synthetic spectra,
at stellar surface,
as in Eq.~(\ref{I_lambda_sg2}).
The monochromatic flux,
at Earth,
is then
\begin{equation}
\Phi_\lambda = {1\over d^2} \sum_i R_i^2 I_{\lambda,i}''\,.
\end{equation}
The monochromatic flux can be compared to spectroscopic observation,
or to narrow-passband photometry,
provided it is converted to the same units
(i.e., calibrated, divided by $\Delta\lambda$),
the object does have too strong spectral features
(within $\Delta\lambda$),
and interstellar extinction is properly taken into account
\citep{Jones_2020ApJS..247...63J}.


\subsection{Implementation notes}

We introduced
two new datasets in Phoebe,
SPE and SED.
The corresponding quantities
\verb|wavelengths|, \verb|fluxes|
were exposed to users (as `twigs').
For example,
\verb|fluxes = b['fluxes@spe01@phoebe01@latest@spe@model'].value|,
or alternatively,
\verb|fluxes = b.get_value('fluxes', dataset='spe01', context='model')|.
Further examples are available as Jupyter notebooks.

Among the grids, which can be used in Phoebe, are
OSTAR \citep{Lanz_2003ApJS..146..417L},
BSTAR \citep{Lanz_2007ApJS..169...83L},
AMBRE \citep{Laverny_2012A&A...544A.126D},
POLLUX \citep{Palacios_2010A&A...516A..13P},
PHOENIX \citep{Husser_2013A&A...553A...6H}, or
POWR \citep{Hainich_2019A&A...621A..85H}.
They can be downloaded from
\url{http://sirrah.troja.mff.cuni.cz/~mira/xitau/}
as text files,
which are automatically converted to binary (npy),
so that next time they are read fast.
In order to use them in Phoebe,
one must prepare a user-defined \verb|gridlist|,
which is a list of files and their associated parameters
($T_{\rm eff}$, $\log g$, $\mathscr{Z}$, or $\mu$).

The sampling of the normalized and absolute grids is 0.01\,\AA\ and 0.1\,\AA, respectively.
Both grids were prepared with the instrumental broadening of this level.
Only the former is suitable for fitting of narrow spectral lines,
while the latter is sufficient in wide spectral lines (like Balmer) and continuum.
It is important that the observed resolution is similar (${\sim}0.01\,{\rm \AA}$),
otherwise narrow spectral lines attributed to individual triangles
would be poorly sampled, which creates spurious numerical noise, or 'wavy' artifacts.
On the other hand,
if the observed resolution is low,
one may optionally apply additional instrumental or rotational broadening
of synthetic spectra during fitting,
which, however, makes it somewhat slower
(see also Fig.~\ref{test_spectroscopy11_fwhm}).

\begin{figure}
\centering
\includegraphics[width=6.5cm]{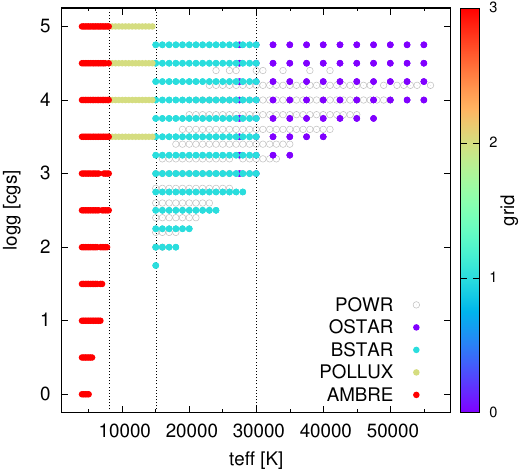}
\caption{
Coverage of the existing grids of normalized synthetic spectra,
in terms of the effective temperature~$T_{\rm eff}$
vs. the gravitational acceleration $\log g$.
The metallicity $\mathscr{Z}$ is solar.
}
\label{gridlist}
\end{figure}

The coverage of the normalized grids is demonstrated in Fig.~\ref{gridlist}.
There are a few regions, where the grids overlap
and one can expect some ``jumps'',
which might be negatively affecting a convergence.
This occurs around 30000\,K,
where OSTAR and BSTAR grids overlap,
and also around 15000 and 8000\,K.
If this problem occurs,
one could comment lines in the \verb|gridlist| to prefer one or the other grid.

Internally, we use Ndpolator%
\footnote{\url{https://github.com/aprsa/ndpolator}}
for linear interpolation or extrapolation in an $n$-dimensional space.
It is as fast as possible (implemented in C),
allowing for a presence of ``voids''
(i.e., a non-uniform grid).
For safety,
the extrapolation method is set to the nearest-neighbor.

\section{Examples} \label{examples}

\subsection{Comparison of complex vs. simplified models}

\begin{figure*}
\centering
\includegraphics[width=9cm]{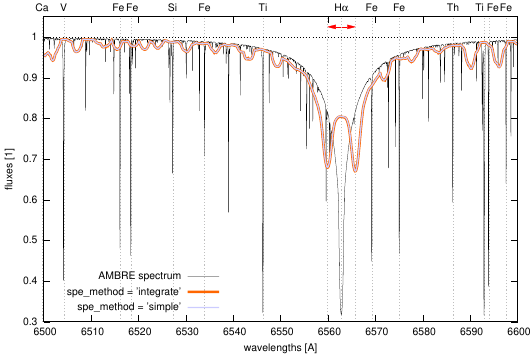}
\caption{
Basic example
of the normalized flux $\Phi_\lambda$ over the wavelength $\lambda$ (in angtr\"oms),
computed for
the default binary in Phoebe
(i.e.,
$P = 1\,{\rm d}$,
$a = 5.3\,R_\odot$,
$m_1 = m_2 \doteq 0.998813\,M_\odot$,
$R_1 = R_2 = 1\,R_\odot$,
$T_1 = T_2 = 6000\,{\rm K}$,
$\log g_1 = \log g_2 \doteq 4.437$).
At the phase 0.25,
the lines of components are shifted
by the radial velocity amplitudes,
$K_1 = K_2 \doteq 134\,{\rm km}\,{\rm s}^{-1}$,
as indicated by the red arrow.
The original, unrotated synthetic spectrum
from the AMBRE grid \citep{Laverny_2012A&A...544A.126D}
is also plotted (black).
Line identifications were taken from
\url{https://bass2000.obspm.fr/solar_spect.php}.
}
\label{test_phoebe9}
\end{figure*}

The first example is a normalized synthetic spectrum $\Phi_\lambda$
obtained for the default binary in Phoebe
(Fig.~\ref{test_phoebe9}).
At the phase 0.25,
the respective spectral lines are double,
but since it is a relatively cool atmosphere (6000\,K)
the numerous metallic lines (Fe, Ti, Th, \dots) overlap
and are thus heavily blended.
Both our models, complex and simplified, agree very well,
because the components are spherical,
the effective temperatures and $\log g$ values are constant,
which implies that all spectra are the same,
only Doppler-shifted and integrated over the surface
(i.e., equivalent to the rotational broadening).

\begin{figure*}
\centering
\includegraphics[width=9cm]{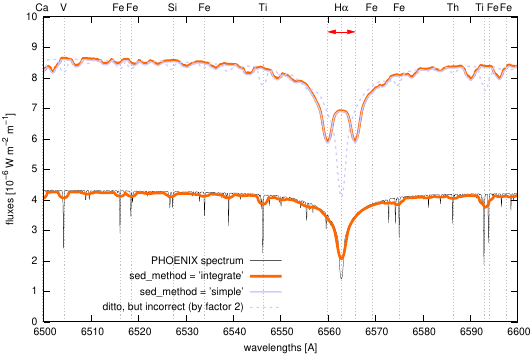}
\caption{
Same as Fig.~\ref{test_phoebe9},
but for the absolute flux $\Phi_\lambda$ (in ${\rm W}\,{\rm m}^{-2}\,{\rm m}^{-1}$)
over the wavelength $\lambda$.
Phases 0 (total eclipse) and 0.25 (out of eclipses) are shown for comparison.
The complex model (\color{orange}orange\color{black})
correctly computes a factor of 2 difference in $\Phi_\lambda$,
but the simplified model (\color{gray}gray\color{black}, dashed)
is unable to account for any eclipses.
The original, unrotated synthetic spectrum
from the PHOENIX grid \citep{Husser_2013A&A...553A...6H}
is also plotted (black).
}
\label{test_phoebe11}
\end{figure*}

In the case of absolute synthetic spectra $\Phi_\lambda$
(in ${\rm W}\,{\rm m}^{-2}\,{\rm m}^{-1}$ units),
the comparison is different
(Fig.~\ref{test_phoebe11}).
The simplified model is slightly offset
with respect to the complex model,
due to slight differences in absolute luminosities
(passband vs. planckian).
However, our models differ at the phase zero,
when the two components are totally eclipsed;
in this case, the simplified model is incorrect
(by a factor of 2).

\subsection{Flux calibration}

\begin{figure*}
\centering
\includegraphics[width=9cm]{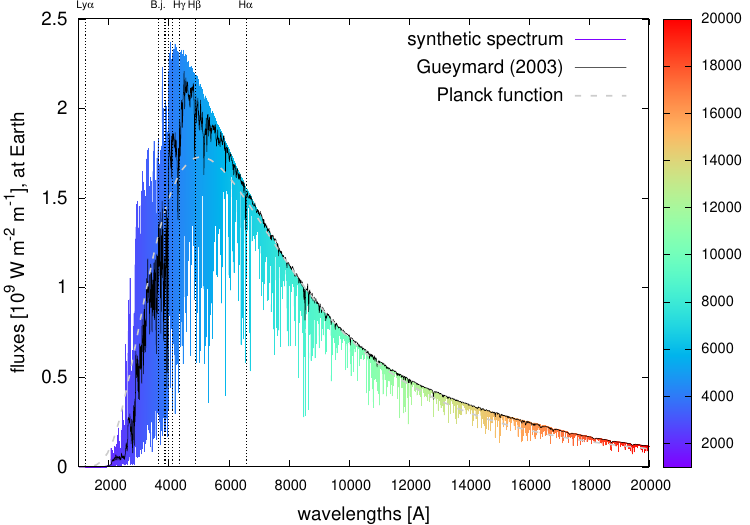}
\caption{
Flux calibration example computed for the Sun at 1\,au distance.
The peak monochromatic flux
$\Phi_\lambda \doteq 2\times10^{-9}\,{\rm W}\,{\rm m}^{-2}\,{\rm m}^{-1}$
is correct.
For comparison,
we plot the synthetic solar spectrum from \cite{Gueymard_2003SoEn...74..355G} (black)
and the Planck function (\color{gray}gray\color{black})
for the effective temperature $T_{\rm eff} = 5770\,{\rm K}$.
}
\label{test_phoebe12}
\end{figure*}

The second example is a flux calibration computed for the Sun at 1\,au distance
(Fig.~\ref{test_phoebe12}).
Its synthetic spectrum was interpolated from the PHOENIX grid;
it agrees well with other solar spectra
\citep{Gueymard_2003SoEn...74..355G}.
Its integral over wavelengths,
$\Phi = \int\Phi_\lambda{\rm d}\lambda \doteq 1338\,{\rm W}\,{\rm m}^{-2}$,
is close to the measured solar constant,
$(1360.8\pm 0.5)\,{\rm W}\,{\rm m}^{-2}$
\citep{Kopp_2011GeoRL..38.1706K}.

\subsection{Limb darkening}

\begin{figure}
\centering
\includegraphics[width=7cm]{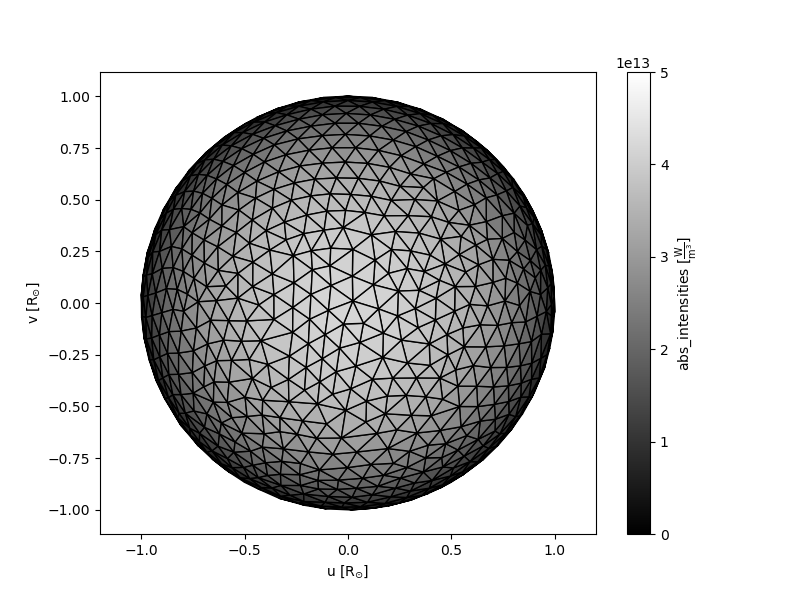}
\includegraphics[width=7.5cm]{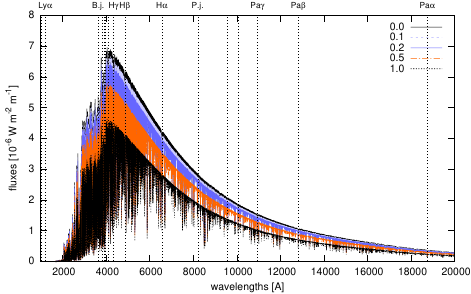}
\caption{
Limb-darkening example
computed for the default star in Phoebe
(i.e.,
$m = 1\,M_\odot$,
$R = 1\,R_\odot$);
observed at a distance of $100\,{\rm pc}$.
Limb darkening coefficients were set to manual, linear;
the values 
0.0,
0.1,
0.2,
0.5,
1.0
were tested.
The SED changes as expected.
}
\label{test_spectroscopy4_limb}
\end{figure}

Limb darkening primarily affects SEDs,
as shown in Fig.~\ref{test_spectroscopy4_limb},
where the decrease of the absolute flux $\Phi_\lambda$
is proportional to the limb darkening coefficient $\epsilon$,
as expected.
Even though this approach is not self-consistent,
because the respective atmospheres also imply the limb darkening
(for $\mu\to 0$),
the full field is not always available.
Moreover, it would require much more extensive, 4-dimensional grids
(for all $\mu$'s).
For reference,
the current grids represent 10 and 14 GB of uncompressed data, respectively
(for $\mu = 1$).
We thus consider our approach to be a good compromise.

\subsection{Rotation}

\begin{figure}
\centering
\includegraphics[width=7cm]{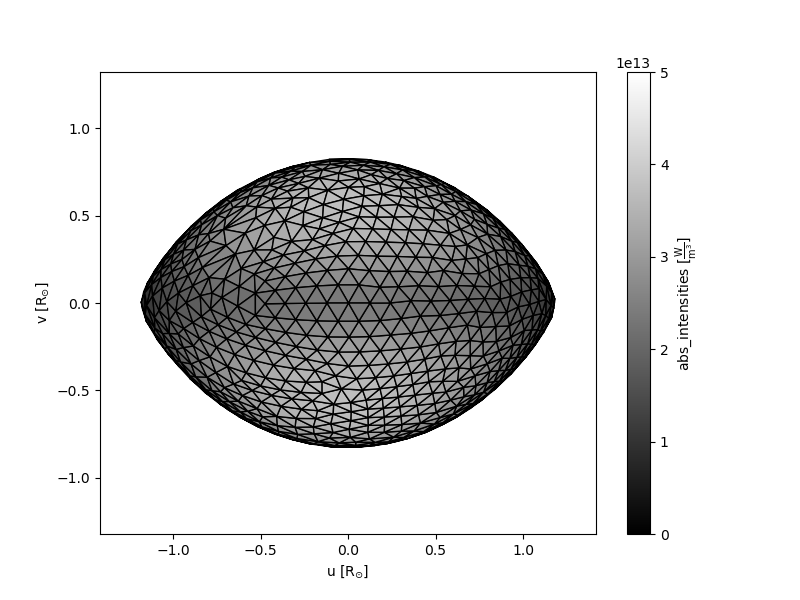}
\includegraphics[width=7.5cm]{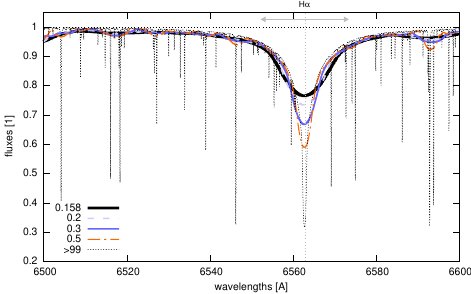}
\caption{
Rotation example
computed for the default star in Phoebe.
The rotation period values
0.16 (close to critical),
0.20,
0.30,
0.50, and
${>}99$~days
were tested.
The spectral lines change as expected.
}
\label{test_spectroscopy6_rotation}
\end{figure}

Rotation affects both SEDs and normalized spectra
(see Fig.~\ref{test_spectroscopy6_rotation}).
Neither the effective temperature~$T_{\rm eff}$,
nor the gravitational acceleration~$\log g$
are constant on the surface
\citep{vonZeipel_1924MNRAS..84..665V},
and spectra must be varied accordingly.
The poles are hotter than the equator,
which leads to observable effects especially in UV,
e.g., for Vega \citep{Aufdenberg_2006ApJ...645..664A}.
In particular, the limb darkening of poles (for $i_\star \to 0$)
must be accounted for,
as it constrains the inclination of the spin axis.
On top of this,
rotation induces Doppler shifts of triangles,
resulting in a rotation broadening of lines,
as expected.

\subsection{Eclipses}

\begin{figure*}
\centering
\begin{tabular}{cc}
\includegraphics[width=6cm]{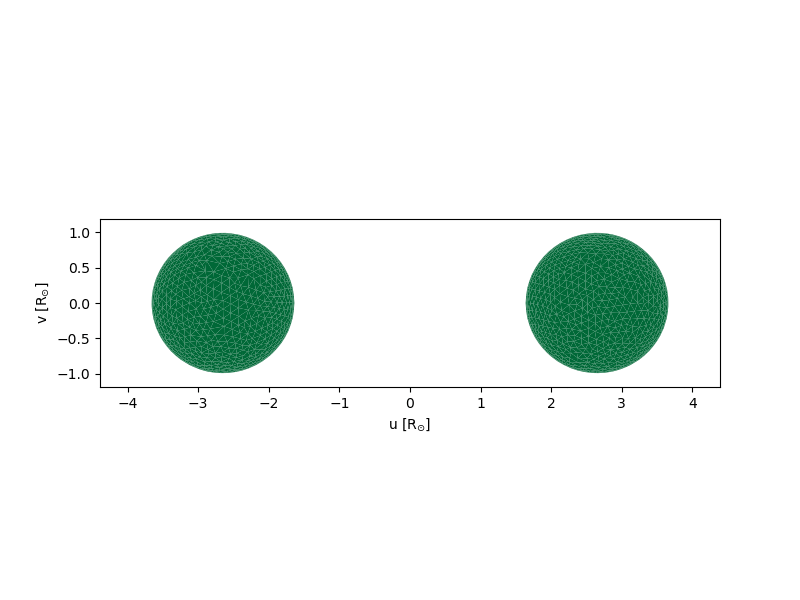} &
\includegraphics[width=6.5cm]{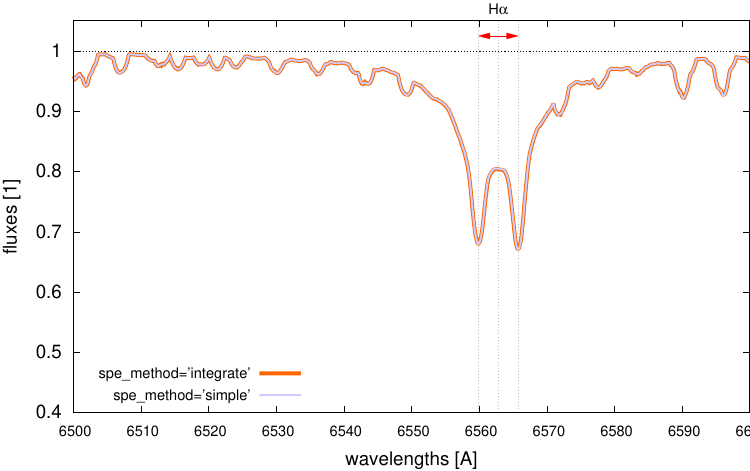} \\[-0.5cm]
\includegraphics[width=6cm]{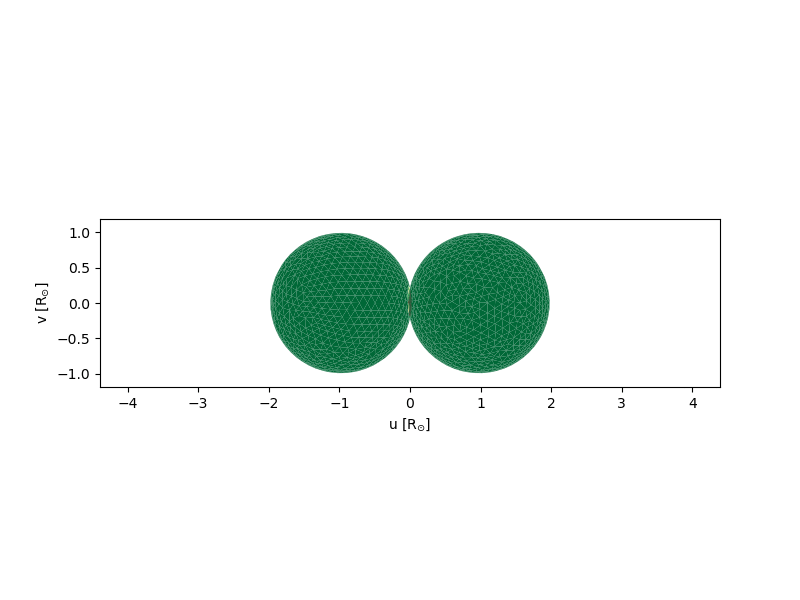} &
\includegraphics[width=6.5cm]{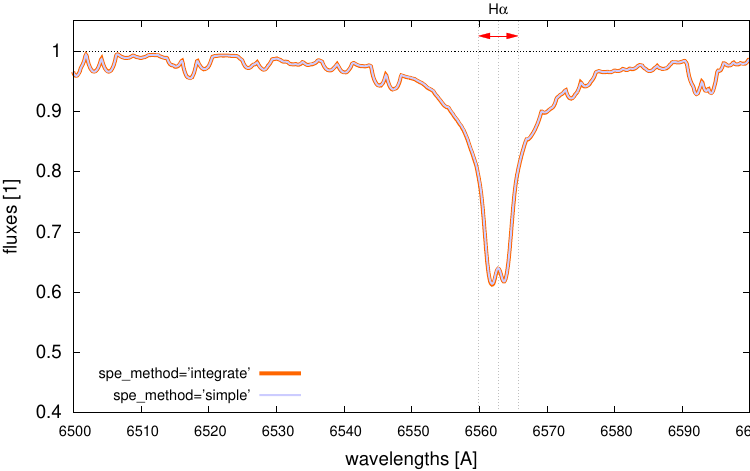} \\[-0.5cm]
\includegraphics[width=6cm]{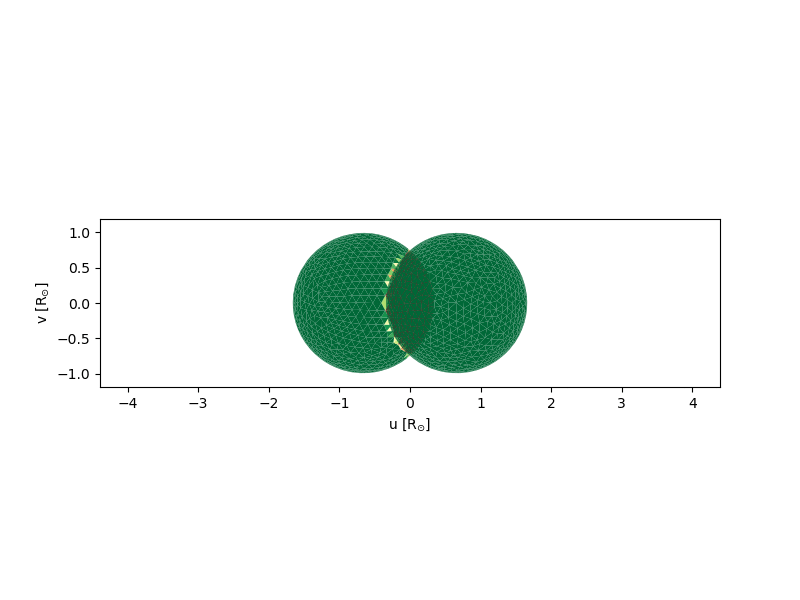} &
\includegraphics[width=6.5cm]{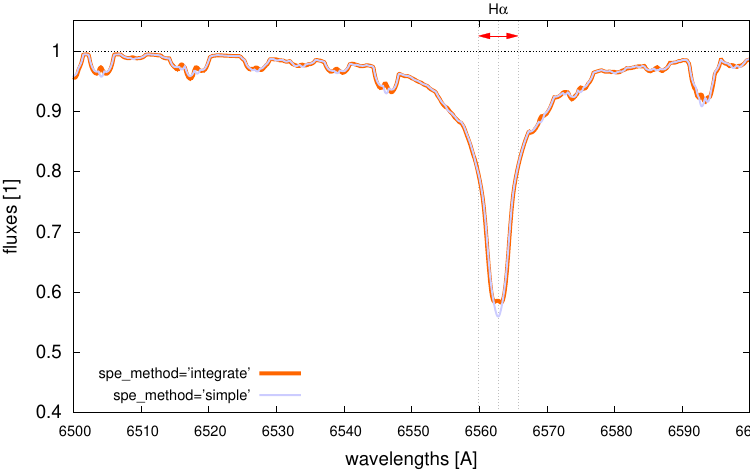} \\[-0.5cm]
\includegraphics[width=6cm]{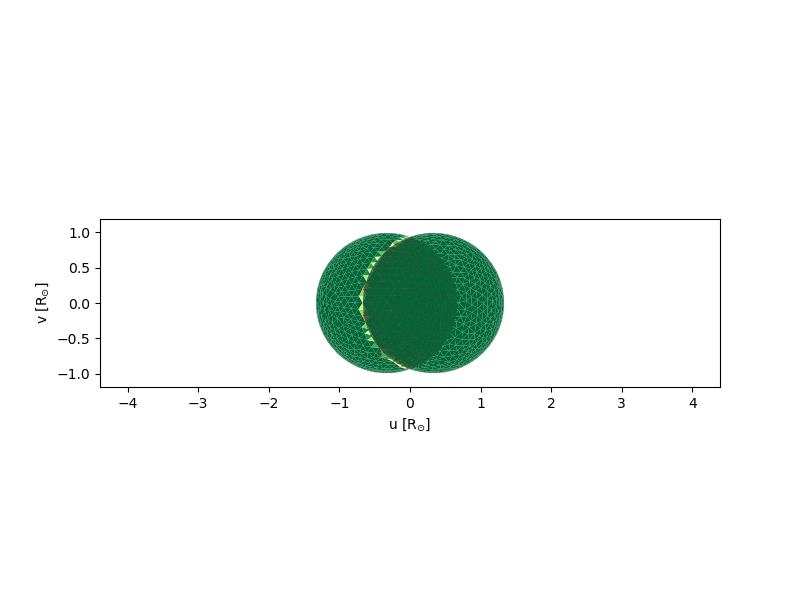} &
\includegraphics[width=6.5cm]{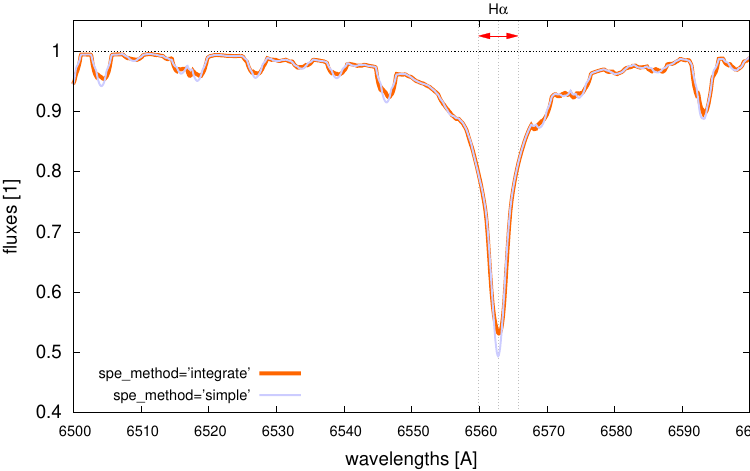} \\[-0.5cm]
\includegraphics[width=6cm]{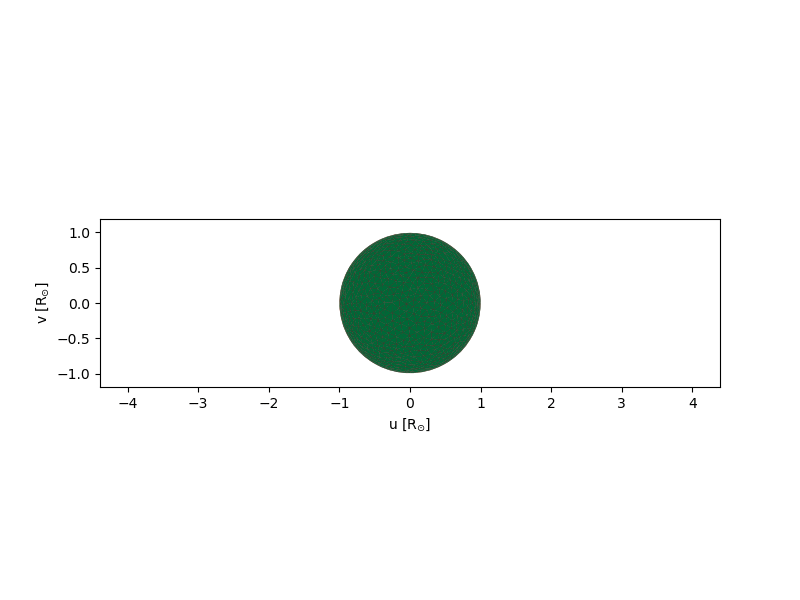} &
\includegraphics[width=6.5cm]{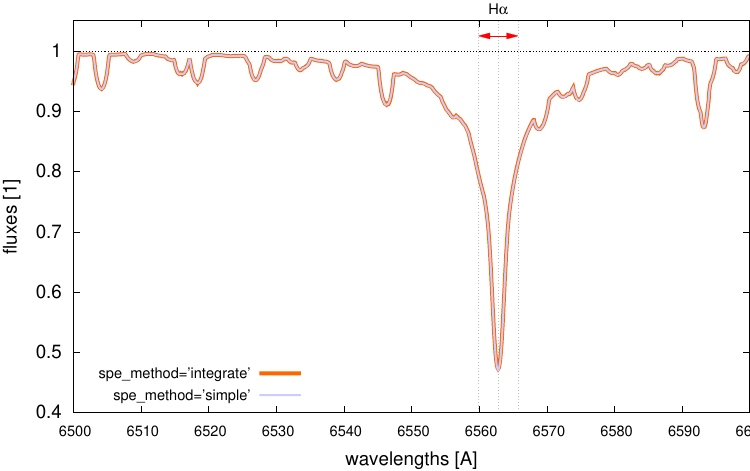} \\
\end{tabular}
\caption{
Eclipse example
computed for the default binary in Phoebe.
Left: Meshes with the fractions of triangles, which are visible.
Right: Normalized synthetic normalized spectra $\Phi_\lambda$
(\color{orange}orange\color{black})
over the wavelength $\lambda$.
During the eclipse, line profiles of the components are blended.
While a simplified model (\color{gray}gray\color{black}) is just a sum of profiles,
the complex model correctly computes asymmetries
arising due to eclipsed parts of the surface.
}
\label{test_spectroscopy3_anim}
\end{figure*}

Finally, we computed normalized synthetic spectra in the course of an eclipse
(Fig.~\ref{test_spectroscopy3_anim}).
The simplified model is again incorrect;
only the complex model shows asymmetries of line profiles
due to the partially eclipsed surfaces.
For the default binary,
this applies to the phases $\phi\in (0; 0.06)$.
One can expect even more pronounced asymmetries
if the respective binary is contact \citep{Abdul_2020A&A...636A..59A},
if one of the components is fast-rotating,
or whenever the temperature difference is substantial.



\section{Conclusions}

We have described the spectroscopic module of Phoebe,
along with a few examples,
how to use it.
Using more datasets (LC, RV, SPE, SED) at the same time
means that models of binaries shall be better constrained,
because spectra and spectral lines contain detailed information
about stellar atmospheres;
more detailed than light curves and radial velocities.

On the other hand,
one should be honest about it,
more datasets sometimes means more problems
--- if systematic errors are present.
Spectroscopy is not a fully independent measurement,
precisely orthogonal to other types of data.
For example, radial velocities suffer from systematics
when lines are blended;
or spectra themselves might be uncertain,
due to their calibration, normalisation, or rectification
(see, e.g., \citealt{Worley_2012A&A...542A..48W,Sacco_2014A&A...565A.113S,Jonsson_2020AJ....160..120J}).

This version of Phoebe also opens the possibility to fit pulsations spectroscopically,
if the radial velocities at the surface are perturbed,
creating waves travelling across line profiles
\citep{Maintz_2003A&A...411..181M,Aerts_2010aste.book.....A}.
This fitting might be problematic though,
if the underlying model for pulsations is insufficient,
e.g., for fast-rotating stars
\citep{Aerts_2023arXiv231108453A}.





\begin{acknowledgements}
This work has been supported by the Czech Science Foundation through
grant 25-16507S (M. Brož).
\end{acknowledgements}

\bibliographystyle{apj}
\bibliography{paper3}


\appendix

\section{Supplementary figures}

\begin{figure}
\centering
\begin{tabular}{ccc}
high resolution, coarse sampling (0.1\,\AA) &
low resolution, fine sampling \\
\includegraphics[width=8cm]{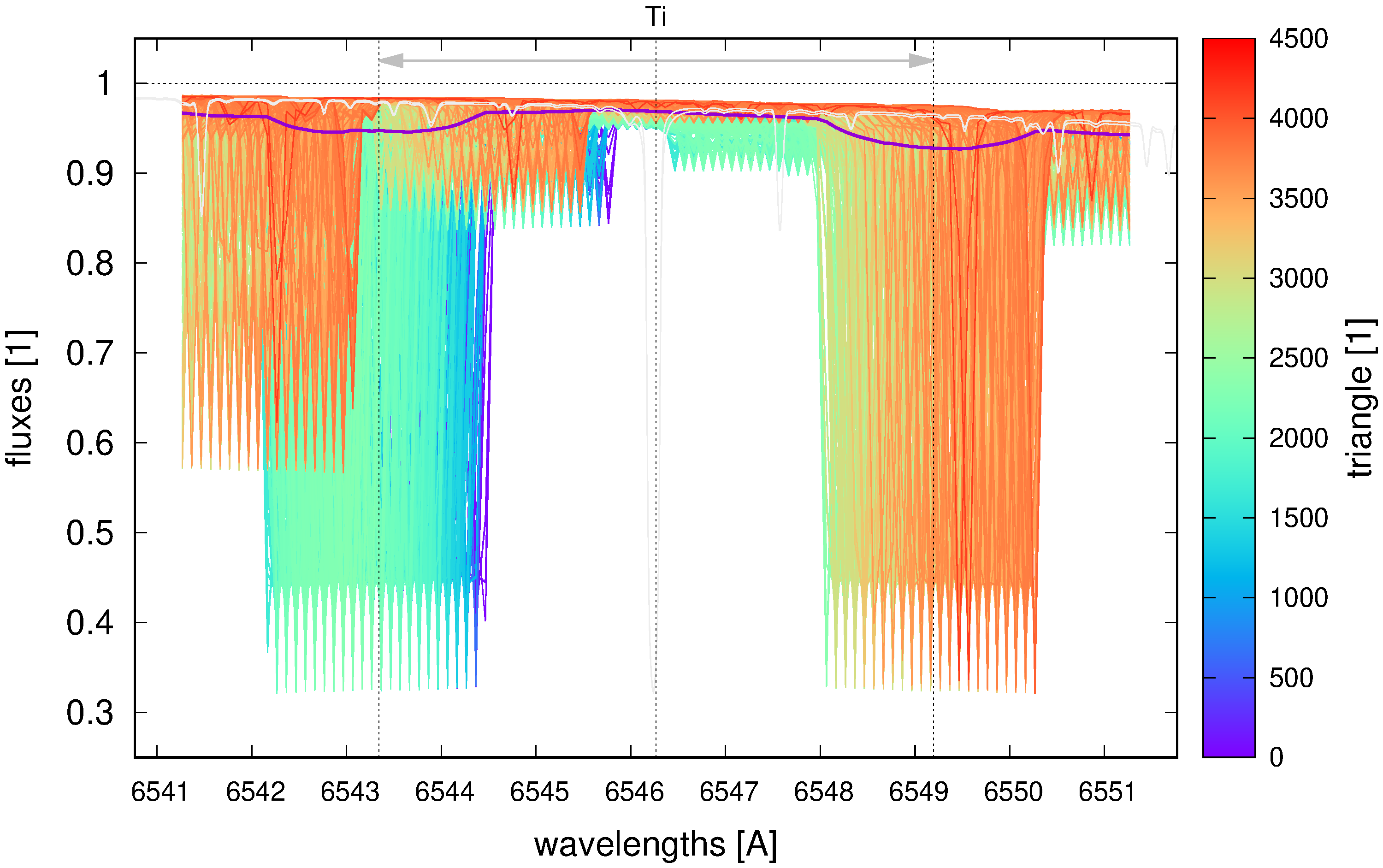} &
\includegraphics[width=8cm]{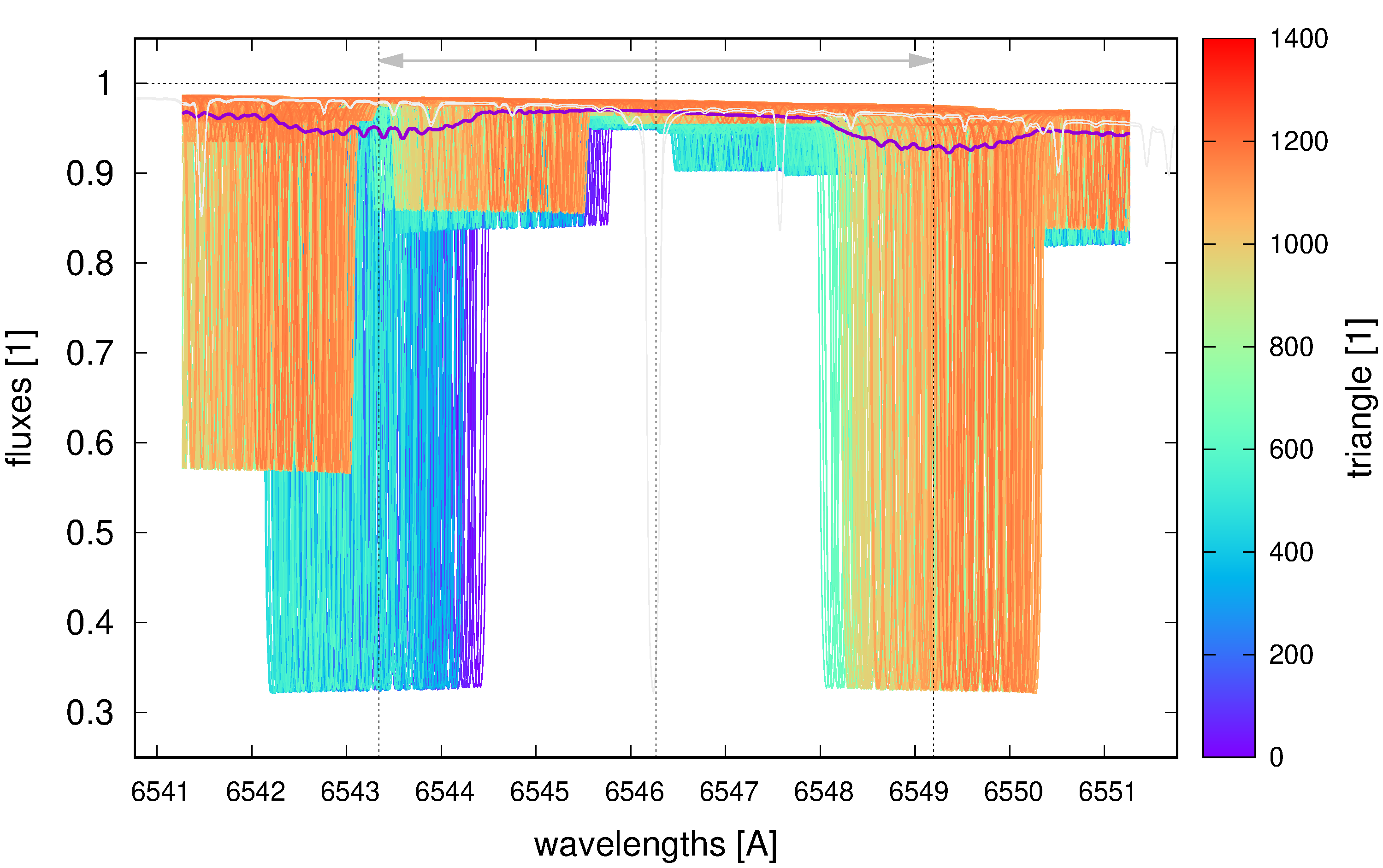} \\
high resolution, fine sampling &
low resolution w. instrumental broadening \\
\includegraphics[width=8cm]{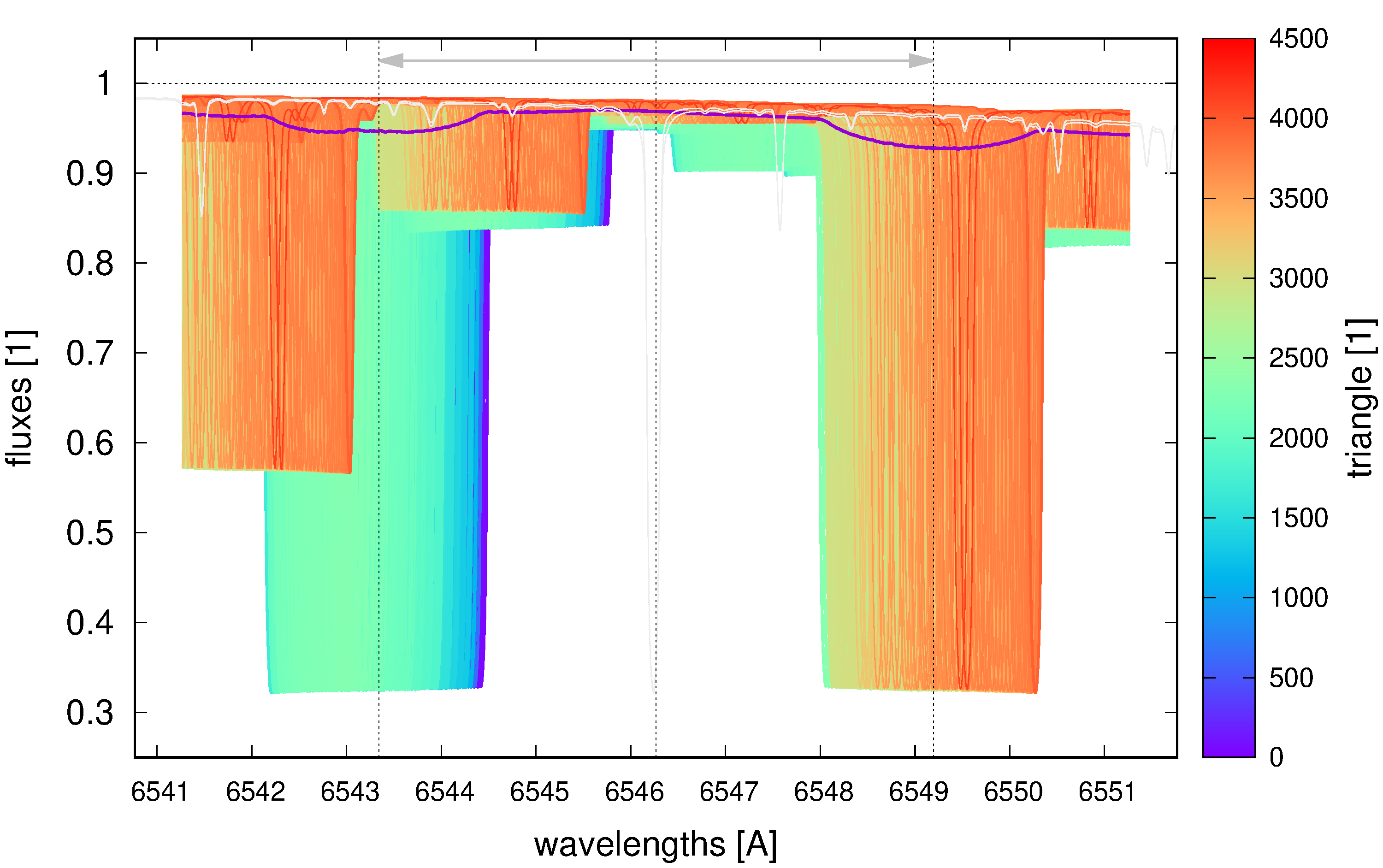} &
\includegraphics[width=8cm]{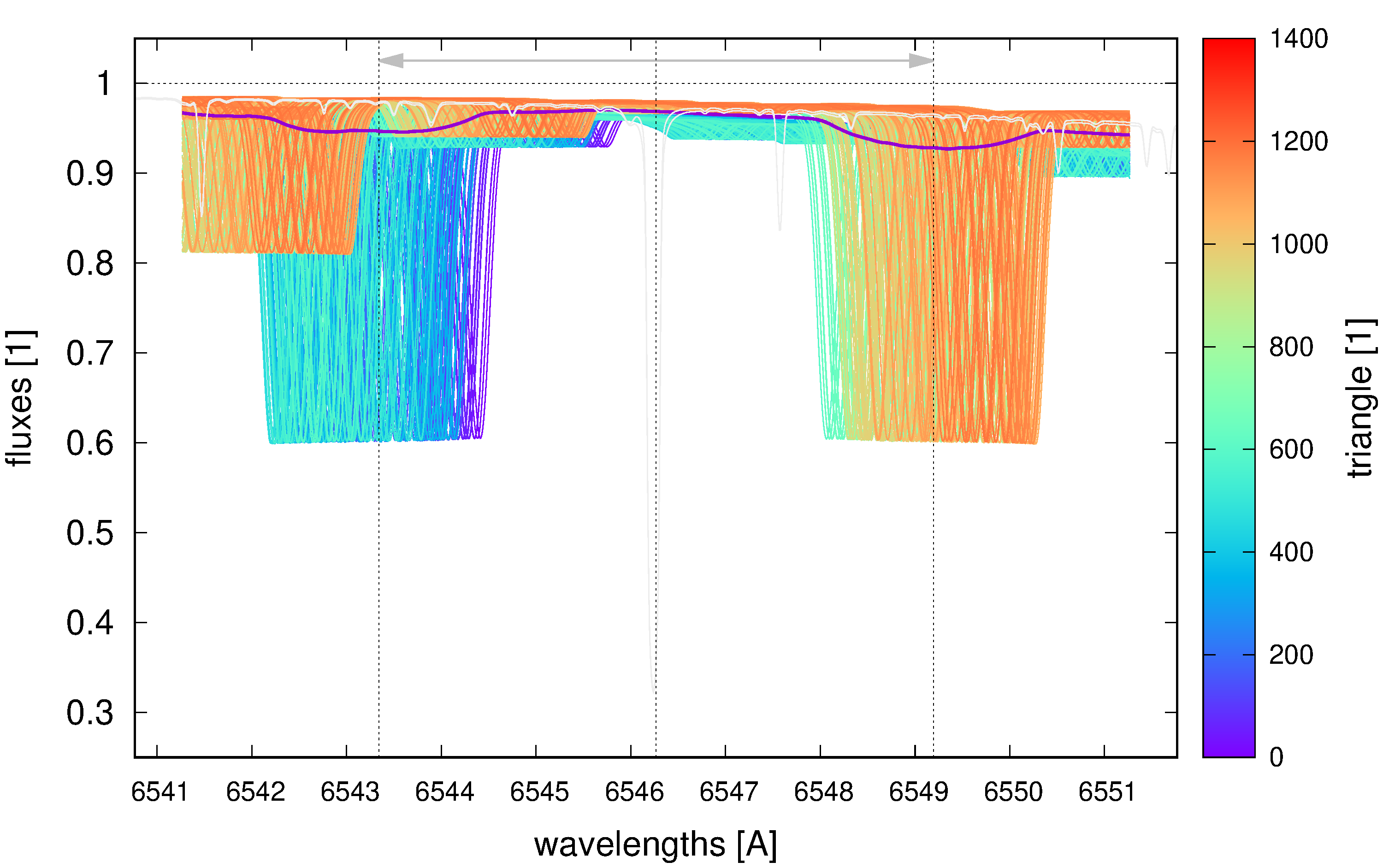} \\
low resolution w. rotational broadening \\[.2cm]
\includegraphics[width=8cm]{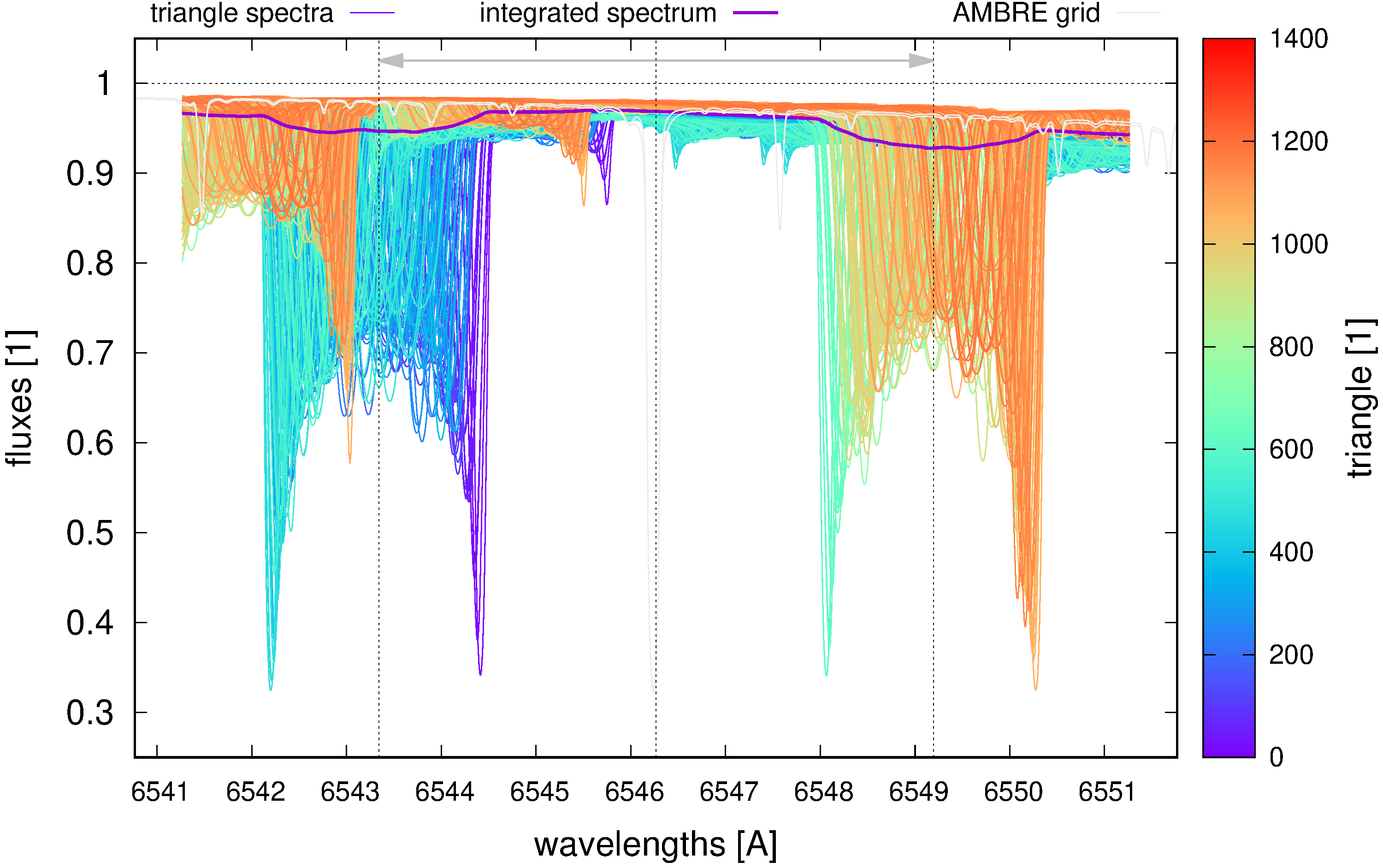} \\
\end{tabular}
\caption{
Synthetic spectra for non-optimal sampling of $\lambda$ or mesh resolution.
Top left:
If the resolution is high (2000 triangles per component),
but the sampling is too coarse (0.1\,\AA),
spectra of individual triangles with narrow lines are poorly sampled,
but the resulting integrated spectrum seems acceptable.
Top right:
If the resolution is too low (500 triangles),
even with fine sampling (0.01\,\AA)
doppler-shifted triangles do not sum up smoothly
and `wave-like' artifacts occur.
Middle left:
One can solve it by using both high resolution and fine sampling.
Middle right:
Or, by using low resolution with instrumental broadening (fwhm = 0.2\,\AA).
Bottom left:
Or, by using low resolution with rotational broadening,
according to a differential radial velocity of each triangle (3 vertices),
which makes spectral lines of individual triangles wide enough,
so that the resulting integrated spectrum seems again acceptable. 
}
\label{test_spectroscopy11_fwhm}
\end{figure}

In Fig.~\ref{test_spectroscopy11_fwhm},
we show how sampling of $\lambda$ and mesh resolution
affect resulting synthetic spectra integrated over triangles.

\end{document}